\pdfoutput=1

\documentclass{article}

\usepackage{microtype}
\usepackage{graphicx}
\usepackage{booktabs} %

\usepackage{hyperref}

\usepackage{xcolor}

\usepackage[accepted]{mlsys2025}
\makeatletter\renewcommand{\Notice@String}{}\makeatother
\hypersetup{pdfauthor={Nathaniel Chen, Cheolsik Byun, SangKyeun Kim, Andrew Rothstein, Ricardo Shousha, Hiro Farre-Kaga, Peter Steiner, Azarakhsh Jalalvand, Egemen Kolemen}}

\usepackage[utf8]{inputenc}
\usepackage[T1]{fontenc}
\usepackage{url,amsmath,amsfonts,array,etoolbox}
\usepackage{tikz}
\usepackage[nolist]{acronym}
\usetikzlibrary{arrows.meta,positioning}

\DeclareRobustCommand{\todo}[1]{\leavevmode{\color{red}[TODO: #1]}}
\newif\ifreviewcolors \reviewcolorsfalse
\ifreviewcolors
  \definecolor{newtext}{RGB}{0,72,186}\definecolor{condtext}{RGB}{0,120,100}
\else
  \definecolor{newtext}{RGB}{0,0,0}\definecolor{condtext}{RGB}{0,0,0}
\fi
\DeclareRobustCommand{\new}[1]{{\color{newtext}#1}}
\newenvironment{revised}{\color{newtext}}{}

\newcolumntype{L}[1]{>{\raggedright\arraybackslash\leavevmode}p{#1}}

\makeatletter
\newcommand{\inputwide}[1]{\begingroup\renewenvironment{table}[1][]{\@dblfloat{table}[t]}{\end@dblfloat}%
  \input{#1}\endgroup}
\makeatother

\newcommand{\appModels}{Appendix~\ref{app:models}}
\newcommand{\appMulti}{Table~\ref{tab:multiall} in Appendix~\ref{app:multi}}
\newcommand{\appSingle}{Table~\ref{tab:single} in Appendix~\ref{app:tables}}

\newif\ifreservespace \reservespacetrue
\ifdefined\noreserve\reservespacefalse\fi
\newcount\reservecount
\newcommand{\reservespace}[1]{\ifreservespace\par\begingroup
  \setlength{\parskip}{0pt plus 0.5pt}\color{red}\reservecount=#1\relax
  \noindent\rule{\columnwidth}{0.4pt}\par
  \loop\ifnum\reservecount>0
    \noindent\vrule height 11pt depth 0pt width 0.4pt\hfill\vrule height 11pt depth 0pt width 0.4pt\par
    \advance\reservecount by -1
  \repeat
  \noindent\rule{\columnwidth}{0.4pt}\par\endgroup\fi}

\newcommand{\PvTbxO}{228}

\newcommand{\PvElmG}{122}
\newcommand{\PvTbxOOneK}{88.8}

\newcommand{\DetOspreadMax}{21.2}

\newcommand{\DetKspreadMax}{29.2}

\newcommand{\DetGspreadMax}{45.2}

\newcommand{\CtKhFourMiss}{19,935}
\newcommand{\CtKhFourN}{20,000}

\newcommand{\CtGhFourMiss}{18,937}

\newcommand{\PrTfFail}{3}

\newcommand{\PrTfModels}{CER, RTCAKENN, TokEye}

\newcommand{\AotCpuWinsA}{7}
\newcommand{\AotXlaBestA}{5}
\newcommand{\AotCpuNA}{9}
\newcommand{\AotCpuWinMaxParamsA}{185k}

\newcommand{\AotGateXla}{3.6}

\newcommand{\AotGateG}{28}

\newcommand{\AotRtcG}{216}

\newcommand{\AotElmXla}{22}

\newcommand{\MmGStreamsFastWorst}{320}

\newcommand{\MmGxStreamsFastMissPct}{8.8}

\newcommand{\MmGxMergedFastWorst}{823}

\newcommand{\MmCnXmixFastWorst}{176}

\newcommand{\MmClkGfourteen}{1,410}

\newcommand{\MmSplitLowClkPct}{1.2}
\newcommand{\MmSplitCerMiss}{112}
\newcommand{\MmSplitCerMissLow}{104}
\newcommand{\MmSplitCerReleases}{36,000}

\newcommand{\MmPredErrMax}{20}
\newcommand{\MmPredErrMed}{5}

\mlsystitlerunning{Is Your AI Fast Enough to Run a Fusion Reactor?}

\begin{document}

\begin{acronym}
    \acro{d3d}[DIII-D]{DIII-D National Fusion Facility}
    \acro{ece}[ECE]{Electron Cyclotron Emission}
    \acro{bes}[BES]{Beam Emission Spectroscopy}
    \acro{fpga}[FPGA]{Field-Programmable Gate Array}
    \acro{cpu}[CPU]{Central Processing Unit}
    \acro{gpu}[GPU]{Graphics Processing Unit}
    \acro{rtml}[RTML]{Real-Time Machine Learning}
    \acro{rtstab}[RTSTAB]{Real-Time Stability}
    \acro{pcs}[PCS]{Plasma Control System}
    \acro{numa}[NUMA]{Non-Uniform Memory Access}
    \acro{sm}[SM]{Streaming Multiprocessor}
    \acro{hpc}[HPC]{High-Performance Computing}
    \acro{mhd}[MHD]{Magnetohydrodynamic}

    \acro{ai}[AI]{Artificial Intelligence}
    \acro{pacman}[PACMAN]{Prediction And Control using MAchiNe learning}
    \acro{cuda}[CUDA]{Compute Unified Device Architecture}
    \acro{cnn}[CNN]{Convolutional Neural Network}
    \acro{xla}[XLA]{Accelerated Linear Algebra}
    \acro{tvm}[TVM]{Tensor Virtual Machine}
    \acro{rtcake}[RTCAKENN]{Real-Time Consistent Automatic Kinetic Equilibrium Neural Network}
    \acro{inpa}[INPA]{Imaging Neutral Particle Analyzer}
    \acro{torbeam}[TORBEAM]{Toroidal Beam Tracing}
    \acro{ml}[ML]{Machine Learning}
    \acro{dnn}[DNN]{Deep Neural Network}
    \acro{onnx}[ONNX]{Open Neural Network Exchange}
    \acro{aot}[AOT]{Ahead-of-Time}
    \acro{nvml}[NVML]{NVIDIA Management Library}
    \acro{tf32}[TF32]{TensorFloat-32}
    \acro{api}[API]{Application Programming Interface}
    \acro{dvfs}[DVFS]{Dynamic Voltage and Frequency Scaling}
    \acro{pca}[PCA]{Principal Component Analysis}
    \acro{cake}[CAKE]{Consistent Automatic Kinetic Equilibrium}

    \acro{elm}[ELM]{Edge-Localized Modes}
    \acroplural{elm}[ELMs]{Edge-Localized Modes}
    \acro{ae}[AE]{Alfv\'en Eigenmodes}
    \acroplural{ae}[AEs]{Alfv\'en Eigenmodes}
    \acro{cer}[CER]{Charge Exchange Recombination} %
    \acro{tm}[TM]{Tearing Mode}
    \acro{itb}[ITB]{Internal Transport Barrier}
    \acro{nbi}[NBI]{Neutral Beam Injection}
    \acro{ech}[ECH]{Electron Cyclotron Heating}
\acro{dct}[DCT]{Discrete Cosine Transform}
\end{acronym}

\twocolumn[
\mlsystitle{Is Your AI Fast Enough to Run a Fusion Reactor?}
\vskip 0.1in

\begin{mlsysauthorlist}
\mlsysauthor{Nathaniel Chen}{pu}
\mlsysauthor{Cheolsik Byun}{pu}
\mlsysauthor{SangKyeun Kim}{pppl}
\mlsysauthor{Andrew Rothstein}{gdm}
\mlsysauthor{Ricardo Shousha}{pppl}
\mlsysauthor{Hiro Farre-Kaga}{pu}
\mlsysauthor{Peter Steiner}{pu}
\mlsysauthor{Azarakhsh Jalalvand}{pu}
\mlsysauthor{Egemen Kolemen}{pu,pppl}
\end{mlsysauthorlist}
\mlsysaffiliation{pu}{Princeton University, Princeton, NJ, USA}
\mlsysaffiliation{pppl}{Princeton Plasma Physics Laboratory, Princeton, NJ, USA}
\mlsysaffiliation{gdm}{Google DeepMind}
\mlsyscorrespondingauthor{Nathaniel Chen}{nathaniel@princeton.edu}
\mlsyscorrespondingauthor{Egemen Kolemen}{ekolemen@princeton.edu}

\mlsyskeywords{Real-time inference, tail latency, GPU, CPU, fusion plasma control}

\vskip 0.3in

\begin{abstract}
In magnetic confinement fusion, compact machine learning algorithms are often integrated into the control loop on CPU-based custom hardware stacks. Deterministic implementations of these algorithms target instabilities which can evolve over microseconds to milliseconds. Staying within this framework may therefore limit the pace of control-system development to the inference time available. With the expansion of hardware acceleration and large models, we aim to bring ML models for fusion control to modern inference engines and compare, based on the physics reactor, the suitability of each engine per model. This study implements a consistent real-time machine learning algorithm for GPUs and optimized CPU packages, including OpenVINO, XLA, and TVM, at DIII-D National Fusion Facility. For the smallest model, one CPU core responds within \AotGateXla\,\textmu s in 99\% of control cycles and an NVIDIA A100 within \AotGateG\,\textmu s. We also evaluate models running simultaneously, as they would run in a typical experiment, and find best use cases per framework.
\end{abstract}
]

\printAffiliationsAndNotice{} %

\section{Introduction}
\label{sec:intro}
To control a magnetic confinement fusion reactor with real-time feedback, \ac{ai} is integrated in the control loop as small models with specialized architectures, so that memory and latency stay within operating constraints. At \ac{d3d} and other fusion facilities, Keras2c is widely used to convert networks into standalone C with static memory allocation, compiled directly into the \ac{pcs}~\citep{conlinKeras2cLibraryConverting2021}. Other models are hard-coded, such as INPA-Net for the \ac{inpa}~\citep{garciaArtificialIntelligencebasedPredictive2025} and a camera-based detachment controller~\citep{chenRegulationCompliantAI2026a}. However, doing so creates a major bottleneck in the development cycle of algorithms for fusion reactors. Recently, \acs{pacman}, a software framework for machine-learning prediction and control, was integrated at \ac{d3d} to address part of this issue by efficiently scheduling multiple models~\citep{rothsteinEnablingIntegratedAI2026}. Despite this, models must remain small enough to fit in the control system's memory.

Understanding the performance constraints of tokamaks requires knowing the timescales of key physical events during a fusion discharge. Tokamak control spans roughly three characteristic timescales. Magnetic control of the plasma's position and shape runs on the main control computers~\citep{kimRealtimePlasmaMonitoring2026} and must act faster than vertical displacement events, which can disrupt the plasma on microsecond timescales ($\tau \sim 10^{-3}$\,ms); kinetic confinement properties, such as temperature and density, evolve over milliseconds ($\tau \sim 1$\,ms)~\citep{rothsteinEnablingIntegratedAI2026}; and the global equilibrium evolves on the order of a second ($\tau \sim 10^{3}$\,ms)~\citep{chenLargeScaleHeterogeneousData2026}. Rapid magnetic stabilization has largely been solved in fusion with classical control, so many of the remaining open problems in feedback control lie in the kinetic regime. These include \acp{ae} driven by fast ions and \ac{tm} precursors, the leading cause of disruptions~\citep{seoAvoidingFusionPlasma2024,chenLargeScaleHeterogeneousData2026}; \acp{elm}, which can be forecast about 8\,ms ahead from 128\,\textmu s of \ac{bes} data, barely in time to apply the coils that suppress them~\citep{joungTokamakEdgeLocalized2024,smithMachineLearningRealtime2025}; and turbulence, whose statistics change over about 10\,ms~\citep{chenLargeScaleHeterogeneousData2026}.

Data from these systems come in different forms and at different rates. The real-time diagnostics that see these dynamics deliver \new{data at MHz rates}. The control system receives magnetics from hundreds of channels, and \ac{ece} from 48 channels, \ac{bes} from 64 channels, and four interferometer chords at 1\,MHz~\citep{kimRealtimePlasmaMonitoring2026}, which is 48{,}000 and 64{,}000 samples per millisecond for \ac{ece} and \ac{bes}. Besides raw signals, it also receives quantities calculated in real time. Slower physics codes are replaced by fast approximations for specific tasks, as inputs to further control steps: \ac{rtcake} approximates the offline \ac{cake} reconstruction in under 8\,ms~\citep{shoushaMachineLearningbasedRealtime2023b}, and TORBEAM-NN, trained on one machine's mirror geometries, replaces a real-time \ac{torbeam} that had itself been simplified~\citep{rothsteinTorbeamNNMachineLearning2025}.

Models implemented on the \ac{cpu} side handle the current loads through clever, targeted engineering and large simplifications that trade inference quality for real-time performance. For example, fast \ac{ece} signals are reduced before an \ac{ae} model sees them: a \ac{cpu} cycle cannot process samples one at a time at 1\,MHz, so the streams are buffered into the cycle~\citep{kimRealtimePlasmaMonitoring2026} and down-sampled to 1\,kHz, even though \acp{ae} oscillate at around 100\,kHz~\citep{jalalvandAlfvenEigenmodeClassification2022}. Reservoir networks are another custom solution, with their own sparse-matrix code, that fit time series with very few parameters~\citep{jalalvandRealTimeAdaptiveReservoir2022}. The current divertor-detachment controller at \ac{d3d} reduces camera images with a linear regression written into the control system instead of a \ac{cnn}~\citep{chenRegulationCompliantAI2026a}. Within the Keras2c framework, the size of a model is also bounded: Keras2c keeps every weight on the stack, which in the \ac{d3d} control system limits a generated model to about 3\,MB and does not scale to many models running together~\citep{rothsteinEnablingIntegratedAI2026}.

Specialized hardware is very fast and can handle large memory bandwidth, but it is difficult to implement and limited to specific inputs and algorithms. For example, \acp{fpga} can evaluate small networks within microseconds, as in particle-physics triggers~\citep{duarteFastInferenceDeep2018}, and \ac{d3d} forecasts \acp{elm} from the full 1\,MHz \ac{bes} stream on one~\citep{daveFPGAAcceleratedRealTimeDiagnostics2026}; superconducting neural-network chips have recently kept up with \ac{d3d} data at a million samples per second~\citep{brookhavenSuperIntelligenceChips2026}. However, machine learning for fusion is still largely research, with new diagnostics and models being added. Thus hardware development for every model is hard to justify.

The obvious solution to these bottlenecks is a \ac{gpu}, but a \ac{gpu} is not automatically faster. In large language models and other machine-learning deployments that tolerate delays of a second, it usually is, because networks are trained and served in batches or minibatches whose express aim is to keep thousands of \ac{gpu} cores busy; a controller instead runs one inference per cycle, at batch size one. A small network then becomes a short series of small kernels, whose time is set by the copies, launches, and synchronization around them rather than by the arithmetic. A good example is \ac{rtcake}, a model with a hundred different small convolutional kernels which can act almost sequentially. Because of these limitations, and the activation barrier of engineering development cycles, \acp{gpu} have entered tokamak control loops only for narrow, fixed computations, more as proofs of concept than for regular use: multi-input, multi-output feedback on HBT-EP~\citep{rathHighspeedMultiinputMultioutput2012} and resistive-wall-mode control on \ac{d3d}, from a \ac{gpu} installed in the \ac{pcs}~\citep{batteySimultaneousStabilizationControl2023}; and the parallel evaluation of a disruption predictor around the operating point, tested on \ac{d3d} data~\citep{boyerActiveDisruptionAvoidance2022}.

Upcoming plasma-state predictors are larger and have longer horizons: networks of millions of parameters for diagnostic analysis~\citep{chen_tokeye_2026}, multimodal foundation models of plasma dynamics~\citep{boschiTokaMindMultiModalTransformer2026}, and generative world models that can take in full \ac{d3d} discharges~\citep{steinerIGNITETokamakWorld2026} cannot fit within the classical paradigm. Improving past models with larger networks is also difficult with past implementations.

In this study, we analyze which control algorithms gain from a \ac{gpu}, and from what size and timescale. \ac{d3d}'s \acs{gpu}-equipped real-time computers allow us to run previously deployed algorithms on the \ac{gpu} and on a \ac{cpu}, with Keras2c, OpenVINO, and the full-network compilers including \ac{xla}~\citep{openxlaXLA2026} and \ac{tvm}~\citep{chenTVMAutomatedEndtoEnd2018}, see which improve, and derive a rough rule, from a task's timescale and a model's parameter count, for what to run on the \ac{gpu}. Our contributions are:
\begin{enumerate}
\setlength{\itemsep}{0pt}
    \item A \ac{gpu} inference path for the two real-time \ac{gpu} stations of the \ac{d3d} \ac{pcs} with TensorRT~\citep{NVIDIATensorRT2026}, and \ac{xla} and \ac{tvm} compiled real-time functions that mirror established implementations of Keras2c.
    \item Latency evaluation on models used in \ac{d3d} operations, ranging from about 3k to 8M parameters for single-model and multi-model inference. This would allow us to estimate the optimal parameter ranges for  real-time deployment.
\end{enumerate}

The mechanisms are known individually in the \ac{gpu} and real-time systems literature~\citep{meiGPUDVFS2017,amertGPUSchedulingTX22017,elliottGPUSync2013}; what is new is their measured effect on the \ac{cpu}-or-\ac{gpu} decision for fusion models on an actual tokamak control system.

\section{Background and Related Work}
\label{sec:background}
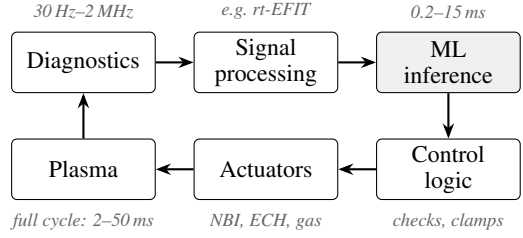
\begin{figure}[t]
  \centering
  \begin{tikzpicture}[
      node distance=6mm and 5mm,
      box/.style={draw, rounded corners=2pt, align=center, minimum height=8mm, minimum width=19mm, inner sep=2pt, font=\footnotesize},
      lbl/.style={font=\scriptsize\itshape, text=gray!120},
      arr/.style={-{Stealth[length=2mm]}, thick}
    ]
    \node[box] (diag) {Diagnostics};
    \node[box, right=of diag] (proc) {Signal\\processing};
    \node[box, right=of proc, fill=gray!12] (ml) {\acs{ml}\\inference};
    \node[box, below=of ml] (ctrl) {Control\\logic};
    \node[box, below=of proc] (act) {Actuators};
    \node[box, below=of diag] (plasma) {Plasma};
    \draw[arr] (diag) -- (proc);
    \draw[arr] (proc) -- (ml);
    \draw[arr] (ml) -- (ctrl);
    \draw[arr] (ctrl) -- (act);
    \draw[arr] (act) -- (plasma);
    \draw[arr] (plasma) -- (diag);
    \node[lbl, above=0.5mm of diag] {30\,Hz--2\,MHz};
    \node[lbl, above=0.5mm of proc] {e.g.\ rt-EFIT};
    \node[lbl, above=0.5mm of ml] {0.2--15\,ms};
    \node[lbl, below=0.5mm of ctrl] {checks, clamps};
    \node[lbl, below=0.5mm of act] {\acs{nbi}, \acs{ech}, gas};
    \node[lbl, below=0.5mm of plasma] {full cycle: 2--50\,ms};
  \end{tikzpicture}
  \caption{In a control cycle, \acs{ml} models run inference on diagnostics or physics quantities, and actuation is issued. The inference budget is what remains of the cycle after the other stages.}
  \label{fig:cycle}
\end{figure}

We can determine rough bounds for our inference latency using physics-informed timescales during a fusion reaction. Since plasma dynamics are chaotic, different tasks may be necessary at rates with different orders of magnitude. Table~\ref{tab:timing} shows documented timing recommendations for control of some events, and Figure~\ref{fig:cycle} shows a single cycle: diagnostics are typically processed into physically interpretable quantities such as a real-time equilibrium \new{(rt-EFIT)}~\citep{laoApplicationMachineLearning2022a}, models consume the resulting signals, and then control logic converts their outputs into commands for actuators such as \ac{nbi}, \ac{ech}, and gas valves.

\begin{table}[t]
  \caption{\acs{d3d} \acs{ml} controllers, their timing constraints, and the reason for each update rate~\citep{rothsteinEnablingIntegratedAI2026,seoAvoidingFusionPlasma2024}. The ITER row is an anticipated requirement~\citep{baylorShatteredPelletDisruption}. \new{$\beta_N$ is the normalized plasma pressure, and ITB an internal transport barrier.}}
  \label{tab:timing}
  \centering\scriptsize
  \setlength{\tabcolsep}{3pt}
  \begin{tabular}{@{}L{0.24\columnwidth}L{0.11\columnwidth}L{0.11\columnwidth}L{0.46\columnwidth}@{}}
    \toprule
    Task & Control cycle & Inference & Reason for the rate \\
    \midrule
    $\beta_N$/\acs{itb} control & 50\,ms & 0.8\,ms & Slower profile evolution; \acs{nbi}/\acs{ech} power \\
    \acs{tm} avoidance & $<$10\,ms & 0.6\,ms & Updates within a $\sim$200\,ms warning horizon; gyrotron steering \\
    \acs{elm} prediction & 2\,ms & 0.2\,ms & Fast edge-event prediction; feeds avoidance logic \\
    \acs{ae} control & $\sim$5\,ms & 0.65\,ms & Fast fluctuation response; \acs{nbi} power \\
    Profile control & 20\,ms & 15\,ms & Profiles evolve over $>$100\,ms; multiple actuators \\
    ITER disruption control & $\sim$15\,ms & $\sim$5\,ms & Anticipated disruption-mitigation response time \\
    \bottomrule
  \end{tabular}
\end{table}

\paragraph{Sharing tasks on a \acs{gpu}.} Shared inference on a \ac{gpu} can be made more predictable in several ways. Clockwork~\citep{gujaratiServingDNNsClockwork2020} relies on the deterministic run time of \ac{dnn} inference to meet latency targets, Paella~\citep{ngPaellaLowlatencyModel2023} orders kernels in software instead of the \ac{gpu}'s hardware scheduler, Orion~\citep{stratiOrionInterferenceawareFinegrained2024} schedules the operators of several clients by their compute and memory needs, and REEF~\citep{hanMicrosecondscalePreemptionConcurrent2022} preempts \new{best-effort} kernels in microseconds so that real-time ones can start. For real-time systems, Fractional GPUs~\citep{jainFractionalGPUsSoftwarebased2019} partitions a \ac{gpu}'s compute and memory bandwidth in software.

\paragraph{Real-time \acsp{gpu}.} \acp{gpu} in real-time systems address many different problems than \acp{gpu} deployed for mass service such as for large language models. For example, GPUSync~\citep{elliottGPUSync2013} addresses real-time locking and budgeting, and \citet{amertGPUSchedulingTX22017} documents how NVIDIA \acp{gpu} schedule work that comes from several streams and processes.

\paragraph{\acs{ml} compilers.} Compilers such as \ac{tvm}~\citep{chenTVMAutomatedEndtoEnd2018} and \ac{xla}~\citep{openxlaXLA2026} are also strong real-time deployment options as they take a model's graph and generate optimized kernels for many back ends. Keras2c~\citep{conlinKeras2cLibraryConverting2021} is a specialized implementation that converts a Keras model into interpretable, dependency-free C code.

\section{Inference Setup}
\label{sec:paths}
We compare five inference methods using only one \ac{cpu} \new{core} and one \ac{gpu} to simulate a real experiment on \new{a} tokamak system.

\begin{table}[t]
  \caption{The two real-time \acs{gpu} stations at \acs{d3d}.}
  \label{tab:stations}
  \centering\scriptsize
  \setlength{\tabcolsep}{3pt}
  \begin{tabular}{@{}L{0.30\columnwidth}L{0.32\columnwidth}L{0.32\columnwidth}@{}}
    \toprule
    & \acs{rtml} & \acs{rtstab} \\
    \midrule
    \acs{cpu} & AMD EPYC 7F52, 16 cores & 4$\times$ Intel Xeon Gold 6154 \\
    \acsp{gpu} & A100-PCIE-40GB & V100-PCIE-16GB, Tesla P4 \\
    Real-time kernel & RedHawk 8.2 & RedHawk 7.5 \\
    TensorRT (\acs{cuda}) & 11.4 (13.3) & 7.0 (9.0), no \acs{cuda} graphs \\
    C compiler & gcc 8.5 & gcc 4.8 \\
    Real-time priority for a user & allowed & refused \\
    \bottomrule
  \end{tabular}
\end{table}

\begin{figure*}[t]
  \centering
  \includegraphics[width=\textwidth]{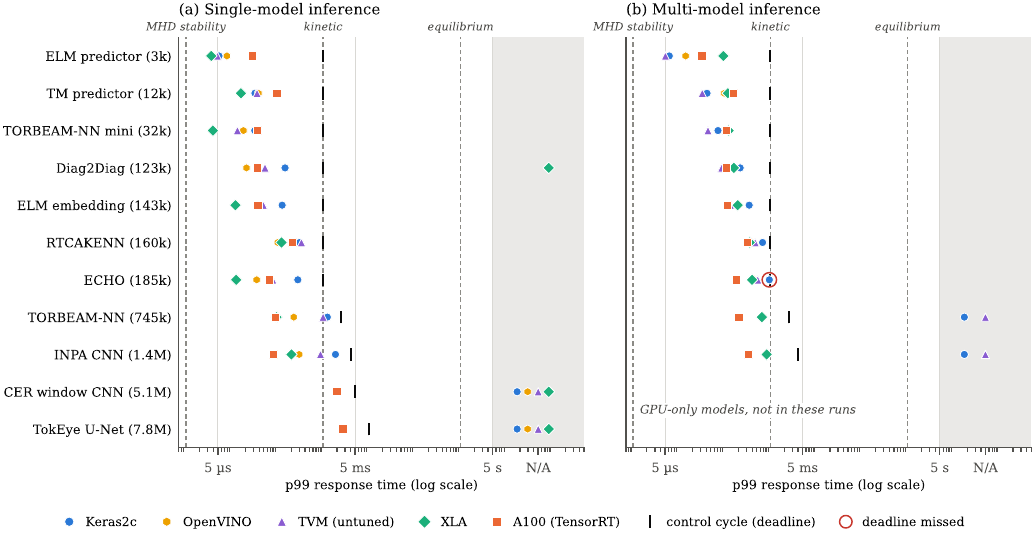}
  \caption{p99 response time of each model on \acs{rtml} with one EPYC core or A100 GPU with real-time priority: (a) each model alone; (b) the models running together, median of three runs. N/A marks a model that a method cannot build or cannot run within its control cycle. Dashed lines mark the physics-relevant magnetohydrodynamic (\acs{mhd}) stability, kinetic, and equilibrium timescales.}
  \label{fig:placement}
\end{figure*}

\begin{table*}[t]
\color{newtext}%
\centering\scriptsize
\setlength{\tabcolsep}{3.5pt}
\caption{p99 response time (\textmu s) on \acs{rtml} with real-time priority, and the slowest response (max) of the same single-model run after its first 20\,ms. Best \acs{cpu} is the fastest of \acs{xla} (X), OpenVINO (O), and untuned \acs{tvm} (T); bold marks the faster of it and the A100. The multi-model columns (medians of three runs) run all models on the A100, with stream priorities or merged engines; all but the \acs{gpu}-only models on one core (\acs{xla} and OpenVINO); and the 1\,kHz models on one core with the rest on the A100. $^\dagger$ marks deadlines missed after a run's first 20\,ms, $^*$ only within it, and --- a model not run.}
\label{tab:main}
\begin{tabular}{@{}lrr rrrrr rrrr@{}}
\toprule
 & & & \multicolumn{5}{c}{Single-model inference} & \multicolumn{4}{c}{Multi-model inference} \\
\cmidrule(lr){4-8}\cmidrule(l){9-12}
Model & Params & Rate & Keras2c & best \acs{cpu} & max & A100 & max & A100, priorities & A100, merged & one core & core + A100 \\
\midrule
ELM predictor & 3k & 1\,kHz & 5.3 & \textbf{3.6\,{\tiny X}} & 5.3 & 28 & 69 & 61 & 823 & 94$^*$ & 3.6 \\
TM predictor & 12k & 1\,kHz & 32 & \textbf{16\,{\tiny X}} & 120 & 97 & 133 & 1,031$^\dagger$ & 823 & 118$^*$ & 19 \\
TORBEAM-NN mini & 32k & 1\,kHz & 32 & \textbf{3.9\,{\tiny X}} & 8.0 & 37 & 111 & 1,106$^\dagger$ & 823 & 123$^*$ & 21 \\
Diag2Diag & 123k & 1\,kHz & 147 & \textbf{21\,{\tiny O}} & 32 & 36 & 77 & 2,121$^\dagger$ & 823 & 160$^*$ & 43 \\
ELM embedding & 143k & 1\,kHz & 127 & \textbf{12\,{\tiny X}} & 17 & 38 & 75 & 2,111$^\dagger$ & 823 & 194$^*$ & 54 \\
RTCAKENN & 160k & 1\,kHz & 306 & \textbf{102\,{\tiny O}} & 119 & 216 & 251 & 973$^\dagger$ & 823 & 354$^*$ & 164 \\
ECHO & 185k & 1\,kHz & 280 & \textbf{13\,{\tiny X}} & 17 & 67 & 116 & 1,126$^\dagger$ & 823 & 398$^*$ & 175 \\
TORBEAM-NN & 745k & 400\,Hz & 1,234 & 96\,{\tiny X} & 173 & \textbf{90} & 137 & 668 & 321 & 651 & 191 \\
INPA CNN & 1.4M & 250\,Hz & 1,857 & 203\,{\tiny X} & 205 & \textbf{82} & 117 & 1,143 & 202 & 831 & 151 \\
CER window CNN & 5.1M & 200\,Hz & --- & --- & --- & 2,008 & 2,073 & 3,120 & 3,058 & --- & 3,398$^\dagger$ \\
TokEye U-Net & 7.8M & 100\,Hz & --- & --- & --- & 2,689 & 2,697 & 3,162 & 3,106 & --- & 4,124 \\
\midrule
\multicolumn{8}{@{}l}{Missed deadlines, all models and runs: after / within the first 20\,ms} & 39,321 / 45 & 0 / 0 & 0 / 27 & 21 / 5 \\
\bottomrule
\end{tabular}
\end{table*}

\paragraph{Compute stations.} \ac{d3d} has two real-time \ac{gpu} stations~\citep{kimRealtimePlasmaMonitoring2026}, and both run RedHawk real-time kernels (Table~\ref{tab:stations}). \ac{rtml} contains an A100, and \ac{rtstab} contains a V100 and a Tesla P4. When running evaluations, we use real-time priority on \ac{rtml} where it is available.

\paragraph{\acs{cpu} with Keras2c and custom methods.} Keras2c is used to write a function essentially capturing a full graph in C for each model; the control system calls it once per cycle. Code generated using Keras2c has the benefit that it allocates nothing at runtime and initializes each weight as a local array, so that each call is debuggable and simply evaluates the layers with plain loops on one core. However, Keras2c supports Keras models only\new{,} so a PyTorch model must be rebuilt in Keras, complicated by the fact that some layers have no Keras2c equivalent. Two of our models are therefore \ac{gpu}-only: the \ac{cer} \ac{cnn} and TokEye. Additionally, some models exceed the memory threshold required for real-time \ac{cpu} deployment even if profiled. For example, \new{the} \ac{inpa} \ac{cnn}'s function alone needs 5.6\,MiB, more than the control system's limit of about 3\,MB~\citep{rothsteinEnablingIntegratedAI2026}. So this was custom built into another workstation.

\paragraph{\acs{cpu} with OpenVINO.} OpenVINO's \ac{cpu} plugin~\citep{OpenvinotoolkitOpenvino2026} reads the same \ac{onnx} graph as \new{TensorRT} and runs it with kernels optimized for the core's instruction set. We run it on a single pinned thread in fp32, so that it uses the same core as existing Keras2c model implementations.

\paragraph{\acs{cpu} with full graph \acs{aot} compilers.} Keras2c and OpenVINO each have drawbacks. Keras2c writes the whole network as one function, but with plain loops instead of optimized kernels, while OpenVINO has optimized kernels but runs the graph through its runtime. \ac{aot} compilers can combine the best of both worlds: \ac{xla}~\citep{openxlaXLA2026} compiles TensorFlow models and \ac{tvm}~\citep{chenTVMAutomatedEndtoEnd2018} compiles \ac{onnx} graphs into a native function with hardwired weights with theoretically much stronger optimization.

\paragraph{\acs{gpu} with TensorRT.} Each model is exported to an \ac{onnx} graph with static shapes and batch size one, which each station's TensorRT (Table~\ref{tab:stations}) compiles. When available, the \ac{cuda} graph of a model is captured in one cycle so that only one launch is needed. Engines are built without \ac{tf32}, for the accuracy reasons given in Section~\ref{sec:gpucost}.

\paragraph{Measurement.} On a synchronous controller, the most important quantity is the response time of each cycle. It is the sum of the timer wake-up error (how late the operating system starts the cycle) and the service time (which includes input copy, kernel launch, \ac{gpu} execution, output copy, synchronization). All runs rely on C or C++ deployment where possible. Additionally\new{,} we use static memory allocation and real-time priority if available for the implementation. Across all evaluations, the inference period uses the same time grid for 20\,s after a warm-up. The rate is 1\,kHz for seven models, 400\,Hz for TORBEAM-NN, 250\,Hz for the \ac{inpa} \ac{cnn}, and 200 and 100\,Hz for the \ac{cer} model and TokEye. \new{Each run records the} timer wake-up error, the service time, and the response time per cycle. Model inputs are pseudo-random values. We compare the p99 of the response time over the whole run. Unlike the single-stream scenario of MLPerf Inference~\citep{reddiMLPerfInferenceBenchmark2020}, which sends each query when the previous one completes and reports the 90th-percentile latency, our queries arrive at a fixed rate with a deadline, so we report the slowest cycle and the missed deadlines. Model descriptions and their sizes are listed in \appModels: a \ac{tm} predictor~\citep{seoAvoidingFusionPlasma2024}, \ac{rtcake}, Diag2Diag~\citep{jalalvandMultimodalSuperresolutionDiscovering2025}, an \ac{elm} predictor and an \ac{elm} embedding~\citep{joungTokamakEdgeLocalized2024,smithMachineLearningRealtime2025}, ECHO~\citep{rothsteinRobustControlECH2026a}, the \ac{inpa} \ac{cnn}~\citep{garciaArtificialIntelligencebasedPredictive2025}, TORBEAM-NN and the smaller TORBEAM-NN mini~\citep{rothsteinTorbeamNNMachineLearning2025}, the \ac{cer} window \ac{cnn}, and TokEye~\citep{chen_tokeye_2026}.

\paragraph{Builds and settings.} On \ac{rtml}, Keras2c and \ac{tvm} are compiled with gcc~8.5 at \texttt{-O3 -march=native}, \ac{xla} with the \texttt{tfcompile} of TensorFlow~2.18 for the EPYC's \texttt{znver2} target, and \ac{tvm}~0.18 with its default, untuned x86 schedules; OpenVINO~2026.4 runs one inference thread. Sections~\ref{sec:det} and~\ref{sec:gpucost} measure what these settings cost.

\section{Single-Model Results}
\label{sec:results}

\subsection{Which Inference Method Is Faster}
\label{sec:which}
Table~\ref{tab:main} and Figure~\ref{fig:placement}a give the p99 response of each model running alone at its rate on \ac{rtml}, and \appSingle\ lists every method on both stations. One \ac{cpu} core is faster than the A100 for \AotCpuWinsA\ of the \AotCpuNA\ models with a \ac{cpu} build, including every model up to \AotCpuWinMaxParamsA\ parameters, and \ac{xla} is the fastest method for \AotXlaBestA\ of them. The \ac{elm} predictor answers within \AotGateXla\,\textmu s with \ac{xla} and \AotGateG\,\textmu s on the A100. OpenVINO is the fastest for Diag2Diag and \ac{rtcake}. The A100 is faster for large models such as the \ac{inpa} \ac{cnn} and TORBEAM-NN. Keras2c, the current method used today, is slower than OpenVINO and \ac{xla} for every model above 30k parameters. With real-time priority, no run missed a deadline.

The methods also scale differently with model size. The A100 has a fixed cost per cycle. For CPU cores, the time grows with the number of parameters, by about 1--2\,ns per parameter with Keras2c and less with \ac{xla} for dense models. This gives an approximate rule for us, where a dense model can stay on a core as long as its time there is below its time on the \ac{gpu} and well inside its budget (Table~\ref{tab:timing}). For example, the whole \ac{elm} pipeline, the embedding and three heads like the predictor, takes \AotElmXla\,\textmu s with \ac{xla} and \PvElmG\,\textmu s on the A100, against a budget of 200\,\textmu s.

\subsection{Determinism and Contention}
\label{sec:det}
With real-time priority, every method on \ac{rtml} is close to deterministic, and the \ac{cpu} methods are the most repeatable. Over the second half of each 1\,kHz run, the slowest cycle is at most \DetKspreadMax\,\textmu s above the run's median for Keras2c, \DetOspreadMax\,\textmu s for OpenVINO, and \DetGspreadMax\,\textmu s on the A100. Like the \ac{gpu} (Section~\ref{sec:split}), the core slows down under light periodic load: OpenVINO's TORBEAM-NN takes \PvTbxOOneK\,\textmu s at 1\,kHz but \PvTbxO\,\textmu s at its own rate of 400\,Hz. With four \ac{cpu}-bound processes on the core, \ac{rtcake} missed \CtKhFourMiss\ of \CtKhFourN\ deadlines with Keras2c and \CtGhFourMiss\ on the A100 (Table~\ref{tab:contention}). However, applying \texttt{SCHED\_FIFO} fixed this.

\begin{table}[t]
\centering\scriptsize
\setlength{\tabcolsep}{4pt}
\caption{\new{\acs{rtcake} at 1\,kHz on \acs{rtml} with busy loops pinned to its core: median and maximum response (\textmu s unless marked) and missed deadlines in 20\,s runs (Keras2c on an idle core with ordinary scheduling: 2\,s). The A100 runs the default TensorRT~11.4 engine.}}
\label{tab:contention}
\begin{tabular}{@{}lrrrr@{}}
\toprule
\new{Load on the core} & \multicolumn{2}{c}{\new{Keras2c}} & \multicolumn{2}{c}{\new{A100}} \\
\cmidrule(lr){2-3}\cmidrule(l){4-5}
 & \new{med / max} & \new{miss} & \new{med / max} & \new{miss} \\
\midrule
\multicolumn{5}{@{}l}{\new{Ordinary scheduling}} \\
\new{idle core} & \new{331 / 335} & \new{0} & \new{226 / 285} & \new{0} \\
\new{1 busy process} & \new{351 / 359} & \new{0} & \new{240 / 292} & \new{0} \\
\new{4 busy processes} & \new{5.3\,s / 10.7\,s} & \new{19,935} & \new{0.1\,s / 0.2\,s} & \new{18,937} \\
\midrule
\multicolumn{5}{@{}l}{\new{\texttt{SCHED\_FIFO} 50}} \\
\new{idle core} & \new{301 / 312} & \new{0} & \new{206 / 250} & \new{0} \\
\new{1 busy process} & \new{302 / 327} & \new{0} & \new{208 / 255} & \new{0} \\
\new{4 busy processes} & \new{308 / 343} & \new{0} & \new{202 / 260} & \new{0} \\
\bottomrule
\end{tabular}
\end{table}

\subsection{Detailed \texorpdfstring{\acs{gpu}}{GPU} Inference Results}
\label{sec:gpucost}
\paragraph{Recipe.} Table~\ref{tab:recipe} removes one part of the A100's recipe at a time for \ac{rtcake}, to see why this model was anomalous and much slower on the A100 than on a \ac{cpu} core in our initial benchmarks. The \ac{cuda} graph and spin-waiting shorten the service time, real-time priority shortens the wake-up, and pinning the thread removes the slowest cycles. TensorRT's default fp32 build uses \ac{tf32} on Ampere \acp{gpu}, and \PrTfFail\ default engines (\PrTfModels) missed an elementwise tolerance of $10^{-4}$ against \ac{onnx} Runtime. Sometimes, driver versions can also be a limiting factor. For example, \ac{rtstab} runs TensorRT~7.0 without \ac{cuda} graphs.

\begin{table}[t]
\centering\scriptsize
\setlength{\tabcolsep}{4pt}
\caption{\new{The A100 breakdown for \acs{rtcake} at 1\,kHz on \acs{rtml}. The thread is pinned to one core with \texttt{SCHED\_FIFO} and locked memory and spin-waits. Times in \textmu s: median wake-up error, median service time, and p99.9 and maximum response. No run missed a deadline.}}
\label{tab:recipe}
\begin{tabular}{@{}lrrrr@{}}
\toprule
\new{Configuration} & \new{Wake} & \new{Service} & \new{p99.9} & \new{Max} \\
\midrule
\new{full recipe} & \new{1.7} & \new{205} & \new{214} & \new{250} \\
\new{without \acs{cuda} graph} & \new{3.0} & \new{295} & \new{326} & \new{601} \\
\new{blocking synchronization} & \new{1.8} & \new{238} & \new{280} & \new{436} \\
\new{without real-time priority} & \new{26.9} & \new{199} & \new{259} & \new{285} \\
\new{\quad with 1\,ns timer slack} & \new{2.3} & \new{199} & \new{224} & \new{254} \\
\new{not pinned to a core} & \new{2.1} & \new{200} & \new{230} & \new{512} \\
\new{on the other \acs{numa} node} & \new{1.9} & \new{200} & \new{214} & \new{252} \\
\new{without locked memory} & \new{2.0} & \new{198} & \new{219} & \new{260} \\
\new{none of these} & \new{45.7} & \new{346} & \new{429} & \new{572} \\
\bottomrule
\end{tabular}
\end{table}

\section{Multi-Model Inference}
\label{sec:multi}
Models run together at rates from 1\,kHz to 100\,Hz. This section asks whether they keep their single-model timing when sharing a core or \ac{gpu}.

\subsection{Executors}
\label{sec:executors}
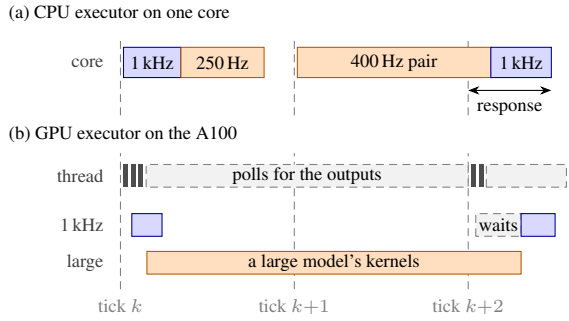
\begin{figure}[t]
  \centering
  \begin{tikzpicture}[x=1mm, y=1mm, font=\scriptsize,
      fast/.style={draw=blue!60!black, fill=blue!15},
      slow/.style={draw=orange!70!black, fill=orange!25},
      wait/.style={draw=gray, densely dashed, fill=gray!10},
      lab/.style={anchor=east, text=black!75}]
    \foreach \x/\t in {16/{tick $k$}, 39/{tick $k{+}1$}, 62/{tick $k{+}2$}} {
      \draw[gray, densely dashed] (\x,19) -- (\x,29.5);
      \draw[gray, densely dashed] (\x,-4) -- (\x,14);
      \node[gray, anchor=north] at (\x,-4) {\t};
    }
    \node[anchor=west] at (0,32.5) {(a) \acs{cpu} executor on one core};
    \node[lab] at (15,26) {core};
    \draw[fast] (16.4,24) rectangle node{1\,kHz} (24,28);
    \draw[slow] (24,24) rectangle node{250\,Hz} (35,28);
    \draw[slow] (39.4,24) rectangle node{400\,Hz pair} (65,28);
    \draw[fast] (65,24) rectangle node{1\,kHz} (73,28);
    \draw[{Stealth[length=1.5mm]}-{Stealth[length=1.5mm]}] (62,22.3) -- (73,22.3);
    \node[anchor=north] at (67.5,21.9) {response};
    \node[anchor=west] at (0,16.5) {(b) \acs{gpu} executor on the A100};
    \node[lab] at (15,11) {thread};
    \foreach \x in {16.4,17.4,18.4} \fill[black!70] (\x,9.5) rectangle ++(0.6,3);
    \draw[wait] (19.4,9.5) rectangle node{polls for the outputs} (62,12.5);
    \foreach \x in {62.4,63.4} \fill[black!70] (\x,9.5) rectangle ++(0.6,3);
    \draw[wait] (64.4,9.5) rectangle (75,12.5);
    \node[lab] at (15,4.5) {1\,kHz};
    \node[lab] at (15,-0.5) {large};
    \draw[fast] (17.5,3) rectangle (21.5,6);
    \draw[slow] (19.5,-2) rectangle node{a large model's kernels} (69,1);
    \draw[wait] (63,3) rectangle node{waits} (69,6);
    \draw[fast] (69,3) rectangle (73.5,6);
  \end{tikzpicture}
  \caption{The two executors over three base ticks (not to scale). (a)~On one core, each model runs to completion, so
  models that run past the next tick delay the 1\,kHz models released there. (b)~On the A100, one thread launches each
  model's \acs{cuda} graph on its own stream and polls for the outputs; kernels of different streams run side by side,
  but a running kernel is not preempted, so a 1\,kHz model's kernels can wait for a large model's.}
  \label{fig:executors}
\end{figure}

We made two executors that run a suite of models from one process, with one real-time thread pinned to one core
(Figure~\ref{fig:executors}). Kernels from different streams share the A100 without preemption. The suites run every network that a deployment would, so the \ac{elm} predictor runs with two more heads of the same size, and TORBEAM-NN with its twin for the other beam polarization. Each configuration ran three times for 20\,s on \ac{rtml}. Figure~\ref{fig:placement}b and Table~\ref{tab:main} show the main runs, and \appMulti\ lists all of
them.

\subsection{Sharing a Core or \texorpdfstring{\acs{gpu}}{GPU}}
\label{sec:sharing}
\paragraph{One \acs{gpu}.} The A100 ran the models that have a \ac{cpu} build without missing a deadline. The slowest 1\,kHz model, \ac{rtcake}, had a p99 of \MmGStreamsFastWorst\,\textmu s, against
\AotRtcG\,\textmu s alone. Running each model in its own process, which would isolate the models, needs NVIDIA's Multi-Process Service: without it, the A100 time-slices between the processes' contexts and nearly every 1\,kHz release missed its deadline, while with it none did. Adding the two \ac{gpu}-only models raised the clock to \MmClkGfourteen\,MHz, but \MmGxStreamsFastMissPct\% of the 1\,kHz releases then missed, and stream priorities did not prevent this. Only merging the models of each rate into one engine met every deadline, with the 1\,kHz engine at \MmGxMergedFastWorst\,\textmu s: 82\% of its cycle, and four times the 200\,\textmu s that Table~\ref{tab:timing} budgets for \ac{elm} prediction.

\paragraph{One \acs{cpu} core.} One EPYC core ran the nine 1\,kHz networks with \ac{xla} and OpenVINO, every p99
within \MmCnXmixFastWorst\,\textmu s. For these, \new{summing the single-model times of the models ahead of each one predicts its p99} within \MmPredErrMed\% \new{for the median model}
and \MmPredErrMax\% at worst. The same core also holds the two TORBEAM-NN networks and the \ac{inpa} \ac{cnn},
missing deadlines only at the first tick, but the 1\,kHz models then slow down (Table~\ref{tab:main}). With Keras2c and \ac{tvm}, TORBEAM-NN alone takes about 1\,ms, a whole 1\,kHz cycle, so one
core runs only the 1\,kHz models, and Keras2c missed a deadline even here.

\subsection{Splitting Between CPU and GPU}
\label{sec:split}
Here, the 1\,kHz models are put on the core and the rest on the A100, \new{where the \ac{cer} model missed \MmSplitCerMiss\ of \MmSplitCerReleases\ releases}. Without the 1\,kHz models, the A100 idled between the large models' bursts, its clock dropped to 765\,MHz in \MmSplitLowClkPct\% of the ticks, and \MmSplitCerMissLow\ of the misses happened at that clock. \new{With every model on the A100, the \ac{cer} model never missed.} On a shared \ac{gpu}, small models are delayed by the large ones but also keep the clock high for them. Placing each model on its fastest hardware therefore does not give the best results.

\section{Limitations}
\label{sec:limits}
Our measurements have a few limits. The optimized \ac{cpu} methods \ac{xla} and \ac{tvm} run without strict tuning, so more threads or tuned schedules may give different results. The inputs are also synthetic random numbers. In the multi-model runs, all models are released together at the first tick; during deployment, these may need to be staggered.

\section{Conclusion}
\label{sec:conclusion}
In this study, we benchmarked the hardware on which the machine-learning models of the \ac{d3d} \ac{pcs} should run directly with the control system. Most models run fastest on a single \ac{cpu} core: with a whole-model compiler such as \ac{xla}, one core is faster than the A100 for every model up to \AotCpuWinMaxParamsA\ parameters. A \ac{gpu} is needed for the multi-million-parameter networks that upcoming controllers will use. However, once several models share the \ac{gpu}, they non-trivially affect each other's scheduling in ways that single-model measurements do not show. A placement that shares a \ac{gpu} therefore has to be tested with all of its models running at their real rates with real-time priority.

\section*{Acknowledgements}
The authors gratefully acknowledge the collaboration of the DIII-D Team supported by the U.S. Department of Energy, Office of Science, Office of Fusion Energy Sciences, using the National Fusion Facility, a Department of Energy Office of Science user facility, under Award No. DE-FC02-04ER54698, DE-AC02-09CH11466, DE-SC0024527, DE-SC0026429, DE-SC0026640. Additional support was provided by the Princeton Laboratory for Artificial Intelligence under Award No. 2025-97.

Disclaimer: This report was prepared as an account of work sponsored by an agency of the United States Government. Neither the United States Government nor any agency thereof, nor any of their employees, makes any warranty, express or implied, or assumes any legal liability or responsibility for the accuracy, completeness, or usefulness of any information, apparatus, product, or process disclosed, or represents that its use would not infringe privately owned rights. Reference herein to any specific commercial product, process, or service by trade name, trademark, manufacturer, or otherwise does not necessarily constitute or imply its endorsement, recommendation, or favoring by the United States Government or any agency thereof. The views and opinions of authors expressed herein do not necessarily state or reflect those of the United States Government or any agency thereof.

\bibliography{main}
\bibliographystyle{mlsys2025}

\appendix
\counterwithin{table}{section}
\numberwithin{figure}{section}  %

\begin{revised}
\section{Models}
\label{app:models}

\begin{table*}[t]
\caption{Models on the stations (\acs{onnx} opset 13, static shapes, batch~1). Parameter counts are of the exported graph and can differ slightly from the source model. All but the \acs{cer} window \acs{cnn} and TokEye also have a Keras2c build; for ECHO, Keras2c builds only the trunk, whose \acs{onnx} subgraph is used for the comparison. In the multi-model runs, the \acs{elm} predictor runs with two more heads of 3,003 parameters each, and TORBEAM-NN with its 743,927-parameter twin for the other beam polarization.}
\label{tab:models}
\centering\scriptsize
\begin{tabular}{@{}L{.24\linewidth}L{.27\linewidth}L{.26\linewidth}r@{}}
\toprule
Model & Inputs & Outputs & Parameters \\
\midrule
\acs{cer} window \acs{cnn} & $48\times39\times256$ preprocessed log spectra & $T_i$, $v_\mathrm{tor}$ mean and std., $48\times39\times2$ each & 5,115,674 \\
TokEye U-Net & $1\times256\times256$ spectrogram & $2\times256\times256$ & 7,846,118 \\
\acs{inpa} \acs{cnn} & $21\times1$ & 21,888 & 1,441,358 \\
TORBEAM-NN & 29 & 1 & 745,232 \\
ECHO & 101; 101; 3; 14 (trunk: 25) & 1; 3; 1; 1 & 185,978 (184,934) \\
\acs{rtcake} & 2; 16; $33\times6$; 15 & $121\times7$ & 159,819 \\
\acs{elm} embedding & 124 & 1,000 & 143,276 \\
Diag2Diag (two checkpoints) & 596 & 596 & 123,412 \\
TORBEAM-NN mini & 24 & 6 & 31,686 \\
\acs{tm} predictor & 11; $33\times5$ & 2 & 11,698 \\
\acs{elm} predictor & 1,000 & 3 & 3,000 \\
\bottomrule
\end{tabular}
\end{table*}

Table~\ref{tab:models} lists the eleven models we compared on the stations. The \ac{aot} compilers cover the models that have a Keras2c build: \ac{tvm} builds all of them, and \ac{xla} all but Diag2Diag since it doesn't have a Keras file. We did not convert TokaMind, because its export needs \texttt{LayerNormalization} (opset~17) and scatter operators that TensorRT~8.0 lacks.

\section{Single-Model Methods}
\label{app:tables}
Table~\ref{tab:single} lists every single-model method on both stations.
\end{revised}

\begin{table*}[t]
\color{newtext}%
\centering\scriptsize
\setlength{\tabcolsep}{3.5pt}
\caption{p99 response time (\textmu s) of each model run alone at its rate: on one \ac{cpu} core with Keras2c, OpenVINO, \ac{tvm} and \ac{xla}, and on each station's \acp{gpu} with TensorRT (11.4 without TF32 on the \ac{rtml} station; 7.0 fp32 on the \ac{rtstab} station). Bold: fastest method of the station; overran: the model takes longer than its period; $^\dagger$: at least one deadline missed. \ac{xla} has no Diag2Diag (a PyTorch model). \ac{rtstab} runs without real-time priority, which the account is refused there.}
\label{tab:single}
\begin{tabular}{@{}lrrrrrrr@{\hspace{8pt}}rrrrrr@{}}
\toprule
 & & & \multicolumn{5}{c}{\ac{rtml}: EPYC 7F52 core / A100} & \multicolumn{6}{c}{\ac{rtstab}: Xeon Gold 6154 core / V100, P4} \\
\cmidrule(lr){4-8}\cmidrule(l){9-14}
Model & Params & Rate & Keras2c & OpenVINO & TVM & XLA & A100 & Keras2c & OpenVINO & TVM & XLA & V100 & P4 \\
\midrule
ELM predictor & 3k & 1\,kHz & 5.3 & 7.9 & 4.9 & \textbf{3.6} & 28 & 41 & 42 & \textbf{16} & 19 & 106 & 68 \\
TM predictor & 12k & 1\,kHz & 32 & 39 & 36 & \textbf{16} & 97 & 181 & 116 & 89 & \textbf{88} & 348 & 460 \\
TORBEAM-NN mini & 32k & 1\,kHz & 32 & 18 & 13 & \textbf{3.9} & 37 & 228 & 50 & 50 & \textbf{24} & 174 & 221 \\
Diag2Diag & 123k & 1\,kHz & 147 & \textbf{21} & 54 & --- & 36 & 666 & \textbf{76} & 112 & --- & 139 & 109 \\
ELM embedding & 143k & 1\,kHz & 127 & 41 & 49 & \textbf{12} & 38 & 709 & 78 & 123 & \textbf{60} & 164 & 125 \\
RTCAKENN & 160k & 1\,kHz & 306 & \textbf{102} & 336 & 123 & 216 & 1,073$^\dagger$ & \textbf{272} & 537 & 562 & 652 & 689 \\
ECHO & 185k & 1\,kHz & 280 & 35 & 79 & \textbf{13} & 67 & 904 & 124 & 114 & \textbf{66} & 265 & 311 \\
TORBEAM-NN & 745k & 400\,Hz & 1,234 & 228 & 997 & 96 & \textbf{90} & overran & 404 & 1,022 & \textbf{258} & 442 & 459 \\
INPA CNN & 1.4M & 250\,Hz & 1,857 & 299 & 876 & 203 & \textbf{82} & overran & 543 & 648 & 427 & 347 & \textbf{310} \\
\bottomrule
\end{tabular}
\end{table*}

\section{Offline Benchmarks}
\label{app:offline}
\begin{revised}
Before the station measurements, we ran an offline benchmark of Keras2c, OpenVINO, and TensorRT on ten networks and model components from \ac{d3d} control and diagnostic pipelines (Table~\ref{tab:bench}): a \ac{tm} predictor~\citep{seoAvoidingFusionPlasma2024}, the kinetic equilibrium model \ac{rtcake}~\citep{shoushaMachineLearningbasedRealtime2023b}, the diagnostic imputation model Diag2Diag~\citep{jalalvandMultimodalSuperresolutionDiscovering2025}, the \ac{elm} predictor~\citep{joungTokamakEdgeLocalized2024,smithMachineLearningRealtime2025}, the electron-cyclotron model ECHO~\citep{rothsteinRobustControlECH2026a}, a \ac{cnn} for \ac{inpa} prediction~\citep{garciaArtificialIntelligencebasedPredictive2025}, TORBEAM-NN and TORBEAM-NN mini~\citep{rothsteinTorbeamNNMachineLearning2025}, the spectrogram model TokEye~\citep{chen_tokeye_2026}, and the multimodal transformer TokaMind~\citep{boschiTokaMindMultiModalTransformer2026}. Keras2c and OpenVINO ran on a Xeon Platinum 8268 \ac{cpu} node and TensorRT~8 on a V100S cluster \ac{gpu}, at batch size one for about 10\,s each.
\end{revised}

\begin{table*}[t]
  \caption{Offline benchmark: mean inference latency in ms on cluster hardware, one $\sim$10\,s run per configuration (Keras2c and OpenVINO on a Xeon Platinum 8268; TensorRT~8 on a V100S). N/A: the runtime does not support the model; ---: not measured. $^\dagger$TokEye's generated C replaces bilinear with nearest-neighbor upsampling, so it is not the same prediction.}
  \label{tab:bench}
  \centering\scriptsize
  \setlength{\tabcolsep}{3pt}
  \begin{tabular}{@{}lrrrrr@{}}
    \toprule
    Model & Parameters & Keras2c & OpenVINO & TensorRT FP16 & TensorRT FP32 \\
    \midrule
    \acs{elm} predictor & 3{,}000 & \textbf{0.00645} & 0.0115 & 0.0435 & 0.0407 \\
    \acs{tm} predictor & 12{,}086 & 0.0470 & \textbf{0.0363} & 0.115 & 0.192 \\
    TORBEAM-NN mini & 32{,}454 & 0.0549 & \textbf{0.0184} & 0.0780 & 0.0740 \\
    Diag2Diag & 123{,}412 & 0.161 & \textbf{0.0218} & 0.0548 & 0.0276 \\
    \acs{rtcake} & 160{,}787 & 0.401 & \textbf{0.135} & 0.240 & 0.236 \\
    ECHO & 186{,}726 & --- & \textbf{0.0527} & 0.211 & 0.146 \\
    TORBEAM-NN & 750{,}040 & 1.42 & 0.188 & 0.0839 & \textbf{0.0725} \\
    \acs{inpa} \acs{cnn} & 1{,}441{,}344 & 2.28 & 0.311 & \textbf{0.0693} & 0.0700 \\
    TokaMind & 6{,}927{,}799 & N/A & 76.9 & --- & \textbf{2.20} \\
    TokEye & 7{,}852{,}002 & 3{,}196$^\dagger$ & 203 & \textbf{1.01} & 1.76 \\
    \bottomrule
  \end{tabular}
\end{table*}

The \ac{elm} model is a linear component of the full \ac{elm} forecasting pipeline since parts of it were installed on an FPGA. The \ac{inpa} model is a \ac{cnn} with 21 inputs and 21,888 outputs; it does not inherit the published reservoir model's accuracy~\citep{garciaArtificialIntelligencebasedPredictive2025}. The \ac{rtcake} graph contains convolutional encoding and decoding layers. TokEye's generated C graph substitutes nearest-neighbor for bilinear upsampling, so its latency is a modified-graph observation. ECHO's C result covers a trunk without its preprocessing. TokaMind timing omits \ac{dct} preprocessing and decoding, and its \ac{gpu} artifact is a constant-folded variant.

\begin{revised}
\section{Multi-Model Methods}
\label{app:multi}
Table~\ref{tab:multiall} lists every configuration of multi-model runs on \ac{rtml} (Section~\ref{sec:multi}). We count a miss within a run's first 20\,ms as part of the cold start, when every model is released at once on a core or \ac{gpu} that has just been idle. Every \ac{cpu} suite with an optimized runtime missed deadlines only during this time.

\begin{table*}[t]
\color{newtext}%
\centering\scriptsize
\caption{Every multi-model configuration on the \ac{rtml} station: the largest p99 response of the 1\,kHz models and of all models (median of the repeats), deadline misses after a run's first 20\,ms and within it (all repeats), releases, and the A100's median streaming-multiprocessor (\acs{sm}) clock. Three 20\,s runs each (Keras2c on four cores: one).}
\label{tab:multiall}
\begin{tabular}{@{}lrrrrrr@{}}
\toprule
Configuration & 1\,kHz p99 (\textmu s) & all p99 (\textmu s) & misses after 20\,ms & first 20\,ms & releases & MHz \\
\midrule
A100, 12 models: one stream & 488 & 628 & 0 & 0 & 603,000 & 930 \\
A100, 12: a stream per model & 320 & 331 & 0 & 0 & 603,000 & 765 \\
A100, 12: streams with rate-monotonic priorities & 319 & 329 & 0 & 0 & 603,000 & 765 \\
A100, 12: one graph per tick, parallel branches & 300 & 303 & 0 & 0 & 603,000 & 765 \\
A100, 12: one graph per tick, in sequence & 642 & 662 & 0 & 0 & 603,000 & 915 \\
A100, 12: merged engines (1\,kHz, 400\,Hz) and INPA & 417 & 417 & 0 & 0 & 603,000 & 825 \\
A100, 12: one process per model & 23.3\,s & 23.3\,s & 540,462 & 580 & 603,000 & 1,410 \\
A100, 12: one process per model, MPS & 288 & 288 & 0 & 0 & 603,000 & 765 \\
A100, 14 models: one stream & 3,303 & 3,942 & 329,623 & 334 & 621,000 & 1,410 \\
A100, 14: a stream per model & 2,230 & 3,256 & 47,583 & 54 & 621,000 & 1,410 \\
A100, 14: streams with priorities & 2,121 & 3,162 & 39,321 & 45 & 621,000 & 1,410 \\
A100, 14: one graph per tick & 3,072 & 3,103 & 281,718 & 282 & 621,000 & 1,410 \\
A100, 14: merged engines with priorities & 823 & 3,106 & 0 & 0 & 621,000 & 1,410 \\
core, 9 at 1\,kHz: Keras2c & 950 & 950 & 1 & 1 & 540,000 & --- \\
core, 9: \acs{tvm} & 529 & 529 & 0 & 0 & 540,000 & --- \\
core, 9: OpenVINO & 263 & 263 & 0 & 0 & 540,000 & --- \\
core, 9: fastest-alone runtime, no \acs{xla} & 237 & 237 & 0 & 0 & 540,000 & --- \\
core, 8 (no Diag2Diag): \acs{xla} & 165 & 165 & 0 & 0 & 480,000 & --- \\
core, 9: \acs{xla}, OpenVINO for Diag2Diag and \acs{rtcake} & 176 & 176 & 0 & 0 & 540,000 & --- \\
core, 12: OpenVINO & 446 & 813 & 0 & 10 & 603,000 & --- \\
core, 12: fastest-alone runtime, no \acs{xla} & 377 & 768 & 0 & 6 & 603,000 & --- \\
core, 12: \acs{xla} and OpenVINO & 398 & 831 & 0 & 27 & 603,000 & --- \\
four cores, 12: Keras2c & 951 & 1,430 & 2 & 2 & 201,000 & --- \\
split, 12: 8 on the core, 4 on the A100 & 207 & 207 & 0 & 0 & 603,000 & 765 \\
split, 14: 8 on the core, 6 on the A100 & 210 & 4,173 & 32 & 4 & 621,000 & 1,410 \\
split, 14: 9 on the core (\acs{xla}), 5 on the A100 & 175 & 4,124 & 21 & 5 & 621,000 & 1,410 \\
split, 14: 12 on the core (\acs{xla}), CER and TokEye on the A100 & 402 & 4,955 & 46 & 42 & 621,000 & 1,410 \\
\bottomrule
\end{tabular}
\end{table*}

\end{revised}

\end{document}